\documentclass[12pt,preprint]{aastex}

\usepackage{graphics,epsfig}
\usepackage{natbib}
\usepackage{lscape}

\def\spose#1{\hbox to 0pt{#1\hss}}
\def\lta{\mathrel{\spose{\lower 3pt\hbox{$\mathchar"218$}}
     \raise 2.0pt\hbox{$\mathchar"13C$}}}
\def\gta{\mathrel{\spose{\lower 3pt\hbox{$\mathchar"218$}}
     \raise 2.0pt\hbox{$\mathchar"13E$}}}

\def\figure#1#2 {\par{\narrower\noindent {\bf Fig. #1}
   \hskip 2mm #2\par}\bigskip\noindent}
\def\table#1#2 {\par{\narrower\noindent {\bf Tab. #1}
   \hskip 2mm #2\par}\bigskip\noindent}

\def\registered{{\ooalign{\hfil\raise .00ex\hbox{\scriptsize R}\hfil\crcr\mathhexbox20D}}}

\usepackage{amsmath}
\usepackage{accents}
\newlength{\dhatheight}

\shorttitle{Climatological and UV-based Habitability}

\shortauthors{Sato, Wang \& Cuntz}

\begin{document}


\title{
Climatological and UV-based Habitability of Possible Exomoons \\
in F-star Systems
}

\bigskip
\bigskip

\author{S. Sato{\thanks{Present address: Life Science Center of Tsukuba Advanced Research
    Alliance, University of Tsukuba, Ibaraki, 305-8577, Japan}}
, Zh. Wang, \and M. Cuntz{\thanks{Corresponding author}}}

\bigskip

\affil{Department of Physics, University of Texas at Arlington, \\
Arlington, TX 76019, USA}
\email{satoko.sato@mavs.uta.edu; zhaopeng.wang@mavs.uta.edu; cuntz@uta.edu}

\bigskip


\begin{abstract}
We explore the astrobiological significance of F-type stars of spectral
type between F5~V and F9.5~V, which possess Jupiter-type planets within
or close to their climatological habitable zones.  These planets, or at
least a subset of those, may also possess rocky exomoons, which potentially
offer habitable environments.  Our work considers eight selected systems.
The Jupiter-type planets in these systems are in notably differing orbits
with eccentricities between 0.08 (about Mars) and 0.72.  We consider the
stellar UV environments provided by the photospheric stellar radiation,
which allows us to compute the UV habitable zones for the systems.
Following previous studies, DNA is taken as a proxy for carbon-based
macromolecules following the paradigm that extraterrestrial biology might
be based on hydrocarbons.  We found that the damage inflicted
on DNA is notably different for the range of systems studied, and also
varies according to the orbit of the Jupiter-type planet, especially
for systems of high ellipticity, as expected. For some systems excessive
values of damage are attained compared to today's Earth or during the
Archean eon.  Considering that the detection of exomoons around different
types of stars will remain challenging in the foreseeable future, we view
our work also as an example and template for investigating the combined
requirements of climatological and UV-based habitability for exosolar objects.
\end{abstract}


\clearpage


\section{Introduction}

The identification of habitable regions around main-sequence stars 
constitutes a vital topic of contemporaneous astrobiological research.
However, previous efforts concern mostly late-type stars at the lower
end of the effective temperature scale \cite[e.g.,][]{lamm13a}, though
notable exceptions include the work on F-type stars by \cite{cock99}
and \cite{bucc06}.  \cite{cock99} studied UV radiation environments
for various main-sequence stars for different types of planetary
atmospheres with different concentrations of N$_2$ and CO$_2$, including
genuine efforts to resemble the Archean Earth.  He concluded that life
may be able to survive in the vicinity of most main-sequence stars of
spectral type F, G and K.  Moreover, \cite{bucc06} introduced the
concept of a UV habitable zone (UV-HZ) for main-sequence stars,
including applications to exosolar systems known at the time of study.

A more recent contribution has been made by \cite{sato14}, henceforth
called Paper~I.  They focused on F-type stars of masses between 1.2 and
1.5~$M_\odot$, and considered numerous important aspects, including
(1) the role of stellar main-sequence evolution, (2) the location
of planets within or relative to the circumstellar habitable zones, and
(3) the general influence of planetary atmospheric attenuation, which was
described through parameterized attenuation functions.  It was found
that the damage inflicted on DNA for planets at Earth-equivalent
(i.e., homeothermic) positions is between a factor of 2.5 and 7.1
higher than for solar-like stars, and
the amount of damage critically depends on the planetary position.
Additionally, there are intricate astrobiological relations for the
time-dependency of damage during stellar evolutionary patterns.  If
atmospheric attenuation is considered, lesser amounts of damage are
obtained in response to the model-dependent attenuation parameters.
Previous work by, e.g., \cite{cock99} suggests that UV radiation
has a momentous potential of inducing significant damage on (or loss of)
hydrocarbon-based life forms, the fundamental question, i.e., ``Is UV
a friend or a foe?", however, still remains unanswered.  Favorable
assessments about the role of UV, also encompassing the origin of life,
stem from numerous studies, as, e.g., the work by \cite{mulk03}.
This study indicates that UV light may have played a key role in the
accumulation of the first polynucleotides through their abiogenic
selection as the most UV-resistant biopolymers, a crucial step for
the origin of self-replicating RNA-type molecules showing sufficient
complexity for undergoing Darwinian evolution.

Setting those fundamental topics aside, noting that they are outside
the main scope of our study, we tend to our main theme, which is
the assessment of the UV environments of F-type stars.  Objects of
interest in such environments include potentially habitable objects
such as Earth-type planets located in the stellar climatological
habitable zone\footnote{The definition of the CLI-HZ follows the
work by \cite{kast93}, which assumes an Earth-like planet with
a CO$_2$/H$_2$O/N$_2$ atmosphere and that habitability requires
the presence of liquid water on the object's surface; see Sect.~2.1
for details, including updates by \cite{kopp13,kopp14}.}
(CLI-HZ). Other possibilities for
the facilitation of habitability include objects associated with
Jupiter-type planets, located in the CLI-HZ, such exosolar Trojan
planets \citep[e.g.,][]{dvor04} or massive exomoons \citep[e.g.,][]{will97}.
Systems with giant planets located inside or crossing into the CLI-HZ
during their orbital motion are typically unable of hosting
habitable planets; see, e.g., \cite{nobl02} and \cite{yeag11} for
results from orbital stability simulations.

Hence, the focus of the present work is on the study of
potential exomoons in the environments of F-type stars possessing
observed Jupiter-type planets.  The systems selected for our study
include HD~8673, HD~169830, HD~33564, $\upsilon$~And, HD~86264,
HD~25171, 30~Ari, and HD~153950 (see Fig.~1 and Table~1) based on
data sets not considering additions due to the {\it Kepler} mission.
Figure 1 conveys the stellar positions in the HR Diagram, indicating
that three stars, i.e., HD~33564, HD~25171, and 30~Ari~B, are in
essence zero-age main-sequence (ZAMS) stars, whereas the other
five stars have started to evolve away from the main-sequence.
Note that the {\it Kepler} mission has revealed many new systems of
F-stars hosting Jupiter-type planets, constituting relevant targets
of future research.

Exomoons, albeit the lack of heretofore detections
have been the topic of numerous previous studies; see, e.g.,
\cite{hell14b} for a recent review on the formation, prospects of
habitability and detection methodologies for exomoons.  Thus, the
continued investigation of exomoons is timely.
The topic of this work is a balanced consideration of the CLI-HZ and
UV-HZ for possible exomoons in F-star systems, with consideration of
Hill stability.  To the best of our knowledge, no such paper has
previously been published along those lines.  It is structured as follows:
In Sect.~2, we describe the theoretical approach, including the key
equations and concepts.  In Sect.~3, we introduce our target star--planet
systems.  In Sect.~4, we address various studies of exomoons, including
the calculation of the Hill's spheres for the exoplanets.  In Sect.~5,
we present the results of our study.  Our summary and conclusions are
given in Sect.~6.


\section{Theoretical approach}

\subsection{Climatological habitable zones}

A crucial aspect in the study of circumstellar habitability is
the introduction of the CLI-HZ, a concept, previously considered by
\cite{kast93} and others.  Even though the centerpiece of our study
is quantifying impact of UV (see Sect. 2.2), the CLI-HZ is important
for two reasons.  First, it conveys reference distances of habitability,
which can be compared to the orbits of the Jupiter-type planets.
Those limits are also relevant to possible exomoons, which are
expected to be in close proximity to their host planets.  Second,
the inner and outer limits of the CLI-HZ can also be used as markers
for the assessment of UV habitability, particularly the quantity
$E_{\rm eff}$ (see Sect. 5.1), thus allowing to interpret our results.

Previously, \cite{kast93} utilized 1-D climate models, which were
state-of-the-art at the time, to estimate the position and width
of the CLI-HZ for a range of main-sequence stars, include the Sun.
The basic premise was the assumption of an Earth-like planet with
a CO$_2$/H$_2$O/N$_2$ atmosphere and, furthermore, that habitability
requires the presence of liquid water on the object's surface.
This work was significantly updated through subsequent studies,
including the work by \cite{kopp13,kopp14}.  For example, it
considers the recent Venus / early Mars (RVEM) limits of the CLI-HZ,
previously considered by \cite{kast93}.

Other important limits for
the CLI-HZs are based on the runaway greenhouse effect (inner limit)
and maximum greenhouse effect (outer limit); in the following, these
limits are used to signify the general habitable zone (GHZ); see Fig.~2.
As described by, e.g., \cite{unde03}, at the inner limit, the
greenhouse phenomenon is enhanced by water vapor, thus promoting
surface warming, which increases the atmospheric water vapor content,
thus further raising the surface temperature.  At sufficient stellar
flux, this will trigger the rapid evaporation of all surface water.
Thus, all water will be lost from the surface, and the water will
also be lost from the upper atmosphere by photodissociation and
subsequent water escape to space.  Concerning the outer limit,
it is assumed that a cloud-free CO$_2$ atmosphere shall still be
able to provide a surface temperature of 273~K.  The most extreme
outer limit is that for the extended habitable zone (EHZ), which
in the solar system extends up to 2.40~AU \citep{misc00}.

For solar-like stars, and assuming an Earth-mass object,
\cite{kopp13,kopp14} identified those limits as approximately
0.95 and 1.68 AU, respectively.  In contrast, \cite{kast93}
identified these values as 0.84 and 1.67~AU, respectively.
Another important limit on habitability is given by the
moist greenhouse limit, which for solar-like stars has been
updated to 0.99~AU.  In the previous work by \cite{kast93}
another limit was identified given by the first CO$_2$
condensation obtained by the onset of formation of CO$_2$ clouds
at a temperature of 273~K, which was however not revisited by
\cite{kopp13,kopp14}.  Nevertheless, in the present work,
the moist greenhouse limit and the limit due to the first
CO$_2$ condensation are used to define the conservative
habitable zone (CHZ), as motivated by a large array of
previous studies\footnote{In the work by \cite{kopp13,kopp14}
the CLI-HZ given by the RVEM limits is referred to as GHZ,
whereas the CLI-HZ between 0.95 and 1.68, as identified for
an Earth-mass planet, is referred to as CHZ.  This notation
is different from that of the present study, which follows
previous convention.}.

The extents of the CHZ, GHZ and EHZ for the target stars
(see Table~1) are given in Table~2.  The results are in
agreement with previous studies, which showed that
the more luminous the star, the
larger the extent of the CLI-HZs.  Even though the CLI-HZs of
F-type stars are considerably wider than those of solar-like
stars, stars with masses larger than 1.2~$M_\odot$ possess
continuously habitable zones for less than the current age
of Earth \citep[e.g.,][]{rush13,sato14,cunt16}.  Nevertheless,
this window of time may still be sufficient for the build-up
of extraterrestrial biospheres.  Furthermore, a higher
stellar luminosity also means that the inner limits of the
respective HZ is placed further outward.

A more thorough
assessment of both the widths and the limits of the various
types of HZs requires the consideration of both the luminosities
and the effective temperatures of the target stars, which is
done as part of our approach\footnote{Another important effect
that is beyond the scope of our study is tidal heating, which
is expected to affect both the inner and outer limits of the
CLI-HZs as previously discussed by \cite{barn08,barn13}, \cite{jack08},
and \cite{lamm14}, among others.}.  For example, it is found that
CLI-HZ of the largest width is obtained for HD~86264, where the
EHZ extends from 1.60 to 5.06~AU.  Somewhat smaller extents
for the CLI-HZ are found for HD~8673 and $\upsilon$~And~A;
here the outer limits of the EHZs are given as 4.37 and 4.39~AU,
respectively.  Relatively small extents for the CLI-HZ are found
for HD~33564 and 30~Ari~B.  Here the inner limits are at
about 0.95~AU, whereas the outer limits extent to nearly 3.0~AU.
These stars are late-type F-stars with luminosities of 1.66 and
1.64 $L_\odot$, respectively (see Table~1).  Figure~3 conveys
the various domains of the CLI-HZs together with the exoplanetary
distances from their host stars, also taking into account the
eccentricities of the stellar orbits.  Detailed observational
results about the various target stars are the focus of Sect.~3.


\subsection{UV-based habitability concepts}

Following Paper~I, we consider the DNA action spectrum as a proxy to simulate
the impact of UV radiation regarding hydrocarbon-based biostructure.
Generally, an action spectrum is the rate of physiological activity plotted
against wavelength or frequency of light. Biological effectiveness
spectra can be derived from spectral data by multiplication with an action
spectrum $S_\lambda(\lambda)$ of a relevant photobiological reaction with the
action spectrum typically given in relative units normalized to unity for, e.g.,
$\lambda = 300$~nm.  The biological effectiveness for a distinct range of
the electromagnetic spectrum such as UV radiation is determined by
\begin{equation}
E_{\rm eff} \ = \
\int_{\lambda_1}^{\lambda_2}E_\lambda(\lambda)S_\lambda(\lambda)\alpha(\lambda)d\lambda \ \ ,
\end{equation}
where $E_\lambda(\lambda)$ denotes the stellar irradiance (ergs~cm$^{-2}$~s$^{-1}$~nm$^{-1}$),
$\lambda$ the wavelength (nm), and $\alpha(\lambda)$ the planetary atmospheric attenuation
function; see \cite{horn95}.  Here $\lambda_1$ and $\lambda_2$ are the limits of integration
which in our computations are set as 200 nm and 400 nm, respectively.  Although a significant
amount of stellar radiation exists beyond 400 nm, this portion of the spectrum is disregarded
in the following owing to the minuscule values for the action spectrum $S_\lambda(\lambda)$
in this regime.  Atmospheric attenuation, also referred to as extinction, results in a  
loss of intensity of the incident stellar radiation.  In Eq.~(1) $\alpha$ = 1 indicates
no loss and $\alpha$ = 0 indicates a complete loss (see Sect. 2.3).

For the computation of the stellar irradiance for targets in circumstellar environments,
typically positioned in the CLI-HZ (see Table~2), the domain of interest in the present study,
a further equation is needed, which is
\begin{equation}
E_\lambda(\lambda) \ \propto \ F_\lambda (R_\ast/d)^2 \ \ .
\end{equation}
Here $F_\lambda$ denotes the stellar radiative flux per unit surface, $R_\ast$ is the
stellar radius and $d$ is the distance between the target and the star.  Note that
Eq.~(2) describes the geometrical dilution of the stellar radiation field. 
Based on previous work, the UV region of the electromagnetic spectrum has been
divided into three bands termed UV-A, UV-B, and UV-C.  The subdivisions are
somewhat arbitrary and differ slightly depending on the discipline
involved.  In Paper~I, we used UV-A, 400-320 nm; UV-B, 320-290 nm; and UV-C,
290-200 nm.  Due to the shape of the action spectrum, it was found that in
the environments of F-type stars, most of the damage occurs with respect to
UV-C.  If the incident stellar radiation is largely blocked in the short
wavelength regime of the DNA action spectrum, due to atmospheric attenuation
(see below), the majority of damage may occur regarding UV-B.  Detailed
information for sets of F-type stars, also considering effects of stellar
evolution, has been described in detail in Paper~I.

Through using an earlier version for the DNA action spectrum
previous results were given by \cite{cock99}.
Action spectra for DNA, also to be viewed as weighting functions, have previously
been utilized to quantify damage due to UV radiation \citep{setl74}.
As discussed in Paper~I, the DNA action spectrum increases by almost four orders
of magnitude between 400 and 300~nm (see Fig.~4).  The reason for this behavior
is the wavelength dependence of the absorption and ionization potential of UV
radiation in this particular regime.  A further significant increase in the DNA
action spectrum occurs between 300 and 200~nm.  Regarding Earth, it is found
that the terrestrial Earth's ozone layer is able to filter out this type of
lethal radiation \citep[e.g.,][]{diff91,cock98}; the terrestrial ozone layer
itself has been an important feature for allowing its oxygen level to rise,
and thus paving the way for the development of more sophisticated life forms.


\subsection{Stellar irradiance}

The treatment of the stellar irradiance follows closely the concept of
Paper~I.  The accurate account of stellar radiation is employed, including 
the adequate spectral energy distribution, by utilizing photospheric models
computed by the PHOENIX code following; see \cite{haus92} and subsequent
work.  The adopted range of models for the F-type stars are in response to
effective temperatures of 6440~K for spectral type F5, 6200~K for F8, and
6050~K for G0 $\equiv$ F10.  Contrary to Paper~I, early-type F stars are
not part of the sample, and their spectral information is thus not needed.
For stars intermediate to that grid, spectral information is interpolated
in accord to the stellar effective temperatures.  The PHOENIX code solves
the equation of state, including a very large number of atoms, ions, and
molecules.  With respect to radiation, the one-dimensional spherically
symmetrical radiative transfer equation for expanding atmospheres is solved,
including a treatment of special relativity.  Opacities are sampled
dynamically over about 80 million lines from atomic transitions and
billions of molecular  lines, in addition to background (i.e., continuous)
opacities, to provide an emerging spectral flux as reliable and realistic 
as possible; see Paper~I for further details and references.

Figure~5 depicts the DNA response to stellar photospheric radiation in
the UV range for selected stellar types between F5 and G2.  The irradiance
has been based on PHOENIX models using smoothing based on a bandbass of
1~nm.  Next, the DNA action spectrum has been multiplied with the smoothed
irradiance data for the range between 200 and 400~nm to attain the data
for the figure.
A relevant ingredient to our study is the consideration of atmospheric
attenuation $\alpha(\lambda)$, which typically results in a notable reduction
of the incident stellar radiation.  Appropriate values for $\alpha(\lambda)$
can be obtained through the analysis of theoretical exoatmospheric models
\citep[e.g.,][ and subsequent work]{seag10} or the usage of model-dependent
historic Earth data \citep[e.g.,][]{cock02}.  Within the scope of the
present work that is focused on the impact of photospheric radiation from
different F-type stars, we consider $\alpha(\lambda)$ as described by a
parameterized attenuation function ATT defined\footnote{${\rm ATT}$ equal
to 0 and 1 indicate the presence and absence of atmospheric attenuation,
respectively; hence, the headers of Fig.~8 and 11 of Paper~I have been
mislabelled and should read ${\rm ATT}=1$ instead.} as
\begin{equation}
{\rm ATT}(\lambda) \ = \ {C \over 2}~\Bigl[ 1 + \tanh (A (\lambda-B)) \Bigr] \ \ .
\end{equation}
Here $A$ denotes the start-of-slope parameter, $B$ (in nm) the center parameter,
and $C$ the maximum (limited to unity) of the distribution; see Fig.~6 and Paper~I
for examples of ${\rm ATT}(\lambda)$.

For example, \cite{cock02} provided information about the ultraviolet irradiance
reaching the surface of Archean Earth for various assumptions about Earth's
atmospheric composition; the latter allow us to constrain the wavelength-dependent
attenuation coefficients.  This particular case is akin to the choice of
$(A, B, C) = (0.02, 250, 0.5)$.  Detailed results for planetary atmospheres
for the build-up and destruction of ozone have been given by, e.g., \cite{segu03}.


\subsection{UV habitable zones}

The concept of UV habitable zones, as considered in our study, closely
follows the previous work by \cite{bucc06}.  It again follows the key idea that,
on the one hand, UV radiation may have provided an important energy source for
early Earth for the synthesis of many biochemical compounds and thus fostering
various biogenesis processes, but, on the other hand, excessive amounts of UV
are destructive (see Sect.~1).  Another consideration includes the ``Principle of
Mediocrity" (PoM), which suggests that the biological system of Earth is about
average, and that the development of life and intelligence outside of Earth is
governed by the same laws and concepts as previously on Earth if proper conditions
are met; see \cite{bucc06} for additional comments and references.  As an example
of exobiology, \cite{fran07} employed the PoM to estimate the maximum number of
habitable planets at the time of Earth's origin in the framework of geobiology.

Akin to Eq.~(1), \cite{bucc06} introduced a measure for the DNA damage due to
UV photons radiated by a star of age $t$ that reaches the top of a planetary or
exomoon atmosphere at a distance $d$ (in AU).  It is expressed as 
\begin{equation}
N^{\star}_{\rm DNA} (d) \ = \
\int_{\lambda_1}^{\lambda_2}S_\lambda(\lambda)\frac{\lambda}{hc}\frac{{\tilde F}(t)(\lambda)}{d^2}d\lambda \ \ ,
\end{equation}
where $S_\lambda(\lambda)$ denotes the DNA action spectrum (see Eq.~1),
${\tilde F}$ the stellar UV flux at 1~AU, $h$ the Planck constant, and $c$
is the speed of light.  Here we use $\lambda_1 = 200$~nm and $\lambda_2 = 400$~nm,
although the choice of 315~nm for $\lambda_2$, as done by \cite{bucc06}, is
acceptable owing to the minuscule values for $S_\lambda(\lambda)$ in the
UV-A regime.  For ${\tilde F}$, the PHOENIX models are adopted, which are
interpolated based on the stellar effective temperature, as needed. 

The domains of the stellar UV-HZs are defined based on the PoM according to
\begin{equation}
N^{\star}_{\rm DNA} (d) \ = \
\eta \ N^{\odot}_{\rm DNA} (1~{\rm AU}) {\vert}_{t=t_{\rm ArcE}^{\odot}}
\end{equation}
with $0.5 \le \eta \le 2.0$, where $\eta = 0.5$ and $\eta = 2.0$ correspond
to the outer and the inner limit of a UV-HZ, respectively\footnote{Note that
the definition of the outer limit of the UV-HZ deviates from that used by
\cite{bucc06}, which does not include the term $S_\lambda(\lambda)$.}.  Moreover,
$N^{\odot}_{\rm DNA} (1~{\rm AU}) {\vert}_{t=t_{\rm ArcE}^{\odot}}$ is determined
based on Eq.~4, using the flux received by the Archean Earth (ArcE), i.e., about
$\sim$3.8~Gyr ago, which is assumed equal to 75\% of the present Sun's radiation
on top of Earth's atmosphere \citep[e.g.,][]{kast88,cock98}.  The limits of the
UV-HZs can be calculated based on Eqs.~(4) and (5) through solving for $d$.
The UV-HZs for our target stars are depicted in Fig.~7; they are also compared
to the CLI-HZs.  It is found that the UV-HZs are always larger than the GHZ, but
are notably smaller than the respective EHZ, in consideration of that they do not
extend as far outward; see Table~2 and 3 for additional information.  Note
that both findings also apply to the Sun, where the UV-HZ extends from 0.71 to
1.90~AU \citep{bucc06}, whereas the EHZ, as defined in the present study, extends
up to 2.40~AU \citep{misc00}.


\section{Target stars and planets}

Next we summarize relevant information about our selection of
star--planet systems considered in the present study, which are: HD~8673,
HD~169830, HD~33564, $\upsilon$~And~A, HD~86264, HD~25171, 30~Ari~B,
and HD~153950 (see Table~1).  The systems consist of an F-type star
and at least one Jupiter-type planet (see Table~4); the latter are
in considerably differing orbits with eccentricities between 0.08
and 0.72.  We also comment on the relationships between the planetary
orbits and the stellar CLI-HZs, subdivided into the CHZs, GHZs, and EHZs
(see Sect.~3.1 and Table~2).  Relevant system properties include the
planetary eccentricities, which are relevant for gauging the prospective
of habitability of possible exomoons.

HD~8673 is an F5~V star\footnote{The spectral types as given have been
determined based on the stellar effective temperatures.  They may thus
be modestly different from those conveyed by the references for the
other stellar properties.} of $T_{\rm eff} = 6413$~K \citep{fuhr08}
with a mass of 1.3~$M_\odot$.  The system's age has been estimated as
2.52~Gyr \citep{take07}. HD~8673 possesses a massive planet or
brown dwarf with a minimum mass of 14.2~$M_{\rm J}$ at $a_{\rm p} =
3.02$~AU \citep{hart10}.  Since HD~8673b has an eccentricity of 0.723,
the star--planet distance varies between 0.84~AU and 5.20~AU.  Hence,
the planet moves beyond both the inner and outer limits of the EHZ.
It also moves beyond both limits of the UV-HZ, with the total extent
of star-planet distance being about a factor of 2 larger than that
of the UV-HZ. 

HD~169830, an F7~V star of $T_{\rm eff} = 6266$~K \citep{nord04} with
a mass of 1.4~$M_\odot$ \citep{fisc05}, is host to two planets with
$a_{\rm p} = 0.81$~AU and 3.6~AU, respectively \citep{mayo04}.  The
planetary minimum masses are 2.88 and 4.04~$M_{\rm J}$, respectively.
The orbit of HD~169830c lies completely within the EHZ, but not within
the UV-HZ, whereas HD~169830b is positioned much closer to the star.
The eccentricity of HD~169830c is given as 0.33.  Therefore, the
star--planet distance varies between 2.41~AU, the periapsis, located
inside the CHZ and 4.79~AU, the apoapsis, which is at a distance
situated between the outer limits of the GHZ and EHZ.  The stellar
age is estimated as 2.3~Gyr \citep{nord04}.

HD~33564 is an F7~V star of $T_{\rm eff} = 6250$~K \citep{acke04} with
a mass of 1.25~$M_\odot$ \citep{nord04}.  It is a relatively inactive
star with an age of 3.0~Gyr \citep{nord04,gall05}.  The star hosts a planet
with a minimum mass of 9.1~$M_{\rm J}$ and a semi-major axis of 1.1~AU.
The planetary orbital eccentricity is given as $e_{\rm p} = 0.34$.
Therefore, the star--planet distance varies between 0.73 and 1.47~AU,
implying that this inner limit is closer to the host star than the
inner limit of the EHZ as well as the inner limit of the UV-HZ,
though this difference is small.  Furthermore, HD~33564 has two stellar
companions \citep{domm02}.  However, they are probably unbound to
the main star as revealed by the high differences between the proper
motions of the components \citep{gall05,roel12}.

$\upsilon$~And is a binary system, consisting of an F8~V star of
$T_{\rm eff} = 6212$~K, $\upsilon$~And~A (HD~9826), and of an M4.5~V
star, $\upsilon$~And~B \citep{lowr02,sant04}.  The separation between
the binary components is given as 750~AU \citep{lowr02}, implying that
any habitable environment of $\upsilon$~And~A would remain unaffected
by $\upsilon$~And~B; see, e.g., the methodological approach by
\cite{cunt14,cunt15} to deduce the magnitude of interdependency in
binary systems.  Radial velocity measurements led to the detection
of four planets around $\upsilon$~And~A, and one of the planets, i.e.,
$\upsilon$~And~Ad, is found to be located within $\upsilon$~And~A's
CLI-HZ and UV-HZ. The semi-major axes of the planets $\upsilon$~And~Ab,
c, d, and e are given as 0.0592, 0.828, 2.51, and 5.25~AU, and their
minimum masses are given as 0.69, 1.98, 4.13, and 1.06~$M_{\rm J}$,
respectively {\citep{curi11}}\footnote{In the work by \cite{curi11},
$\upsilon$~And~Ac and d are exchanged; however, in this study we
use the notation given by the sequence of observation.}.
The eccentricity of $\upsilon$~And~Ad
is identified as 0.299 \citep{curi11}, and its unprojected mass has
been estimated as 10.19~$M_{\rm J}$ \citep{barn11}.  Its periapsis is
at 1.76~AU, a distance close to the inner limit of the GHZ.  The planet
stays inside the EHZ at its apoapsis at 3.26~AU.  This system is also
known for the mutual inclination between the planets c and d, which
is as large as 30$^{\circ}$ \citep{barn11}.

HD~86264 is an F8~V star of $T_{\rm eff} = 6210$~K with
a mass of 1.42~$M_\odot$ \citep{fisc09}.  The star possesses a planet
with a minimum mass of 7.0~$M_{\rm J}$ at $a_{\rm p} = 2.86$~AU, i.e.,
slightly beyond the outer limit of the CHZ.  Since the planet has an
eccentricity of $e_{\rm p} = 0.7$, it approaches the star at about half
the distance of the inner limit of the EHZ at its periapsis, and approaches
the outer limit of the EHZ at its apoapsis; the corresponding distance
range is between 0.86~AU and 4.86~AU.  The planet also crosses the
inner and outer limits of the UV-HZ.  HD~86264 has an age of 2.24~Gyr,
and it is moderately active \citep{fisc09}.

HD~25171 is an F9~V star of $T_{\rm eff} = 6160$~K with a mass of
1.09~$M_\odot$ \citep{mout11}.  It is a non-active star with an
age of 4.0~Gyr.  It is also host to a Jupiter-type planet with
a minimum mass of 0.95~$M_{\rm J}$ in a nearly circular orbit with
an eccentricity of $e_{\rm p} = 0.08$, i.e., about the value of
Mars.  The planet orbits the star close to the outer edge of the EHZ,
and most of the planetary orbit also lies beyond the stellar UV-HZ.
The planet's semi-major axis is identified as 3.02~AU.  Hence, the
periapsis and apoapsis are given as 2.78~AU and 3.26~AU, respectively.
The outer limit of the EHZ is identified as 3.13~AU, thus nearly
coinciding with the planet's apoapsis.

30~Ari is a triple star system, consisting of the single line spectroscopic
binary, 30~Ari~A (HD~16246), and the single star, 30~Ari~B (HD~16232).
30~Ari~A (main component) is an F3~V star of $T_{\rm eff} = 6668$~K,
and 30~Ari~B is an F9~V star of $T_{\rm eff} = 6152$~K \citep{nord04}.
The separation between 30~Ari~A and B is about 1520~AU
\citep{perr97,zomb07,guen09}, which ensures that any habitable environment
of 30~Ari~A and B would not interfere.  30~Ari~B hosts a planet at
$a_{\rm p} = 0.995$~AU.  This distance is slightly farther from the star
than the inner limit of the EHZ, which is at 0.947~AU.  Furthermore,
half of the planetary orbit is closer to the star than the inner limit
of the UV-HZ.  The periapsis is at 0.71~AU, but a large part of the
orbit is in the CLI-HZ. The apoapsis is at 1.28~AU, located in the
CHZ based on an orbital eccentricity of 0.289.  The stellar mass and
the minimum mass of the planet are given as 1.11~$M_\odot$ \citep{nord04}
and 9.88~$M_{\rm J}$ \citep{guen09}, respectively.  The age is estimated
as 0.91~Gyr \citep{guen09}, implying that it is a very young star.

HD~153950 is an F9.5~V star with an effective temperature of
$T_{\rm eff} = 6076$~K \citep{mout09}, thus representing the
low temperature limit of our sample.  HD~153950 is host to a planet
with a minimum mass of 2.73~$M_{\rm J}$ and a semi-major axis of
$a_{\rm p} = 1.28$~AU.  The planet's semi-major axis corresponds
to a distance about halfway between the inner limit of the EHZ and
the inner limit of the GHZ.  Moreover, similar to the case of 30~Ari,
about half of the planetary orbit is closer to the star than the inner
limit of the UV-HZ.  At its apoapsis, the planet is located in the CHZ,
but at the periapsis, it approaches the star as close as 0.84~AU. The
eccentricity is $e_{\rm p} = 0.34$ \citep{mout09}.  The host star has
a mass of 1.12~$M_\odot$, and its age is estimated as 4.3~Gyr.


\section{Studies of exomoons}

Exomoons, despite that there are no detections yet, have been the topic of
numerous studies.  Examples include the work by \cite{will97}, \cite{kipp09},
\cite{kalt10}, \cite{barn02}, \cite{hell13a}, \cite{cunt13}, \cite{hell13b},
and \cite{hell15}.  They consider a large variety of aspects including
orbital stability considerations, formation of exomoons around super-Jovian
planets, magnetic shielding of exomoons beyond the circumplanetary habitable
edge, deciphering spectral fingerprints (which could potentially lead to the
detection of life), and the lower mass limit of exomoons required for retaining
a substantial and long-lived atmosphere.

Strictly speaking, exomoons could
also be akin to terrestrial or super-Earth planets in size and mass if orbiting
a Jupiter-type planet.  In this case, the existence and retention of the moon's
armosphere is expected to be virtually guaranteed; see, e.g., \cite{tsia16} for
a recent observational study.  Atmospheres of large exomoons might also enjoy
the benefit of magnetic protection \citep[e.g.,][ and subsequent work]{lamm03,grie04},
which will be able to counteract the impact of energetic stellar radiation,
especially for young stars \citep[e.g.,][ and references therein]{fran14,cunt16}.
Detailed studies about atmospheric losses due to energetic stellar radiation
have previously been given by \cite{erka13}, \cite{lamm13b}, and \cite{cohe15},
among others.

Additional studies about exomoons have been devoted to search methodologies as
discussed by, e.g., \cite{awip13}, \cite{kipp13}, \cite{kipp14}, \cite{hell14b},
and \cite{noyo14}. The work of \citeauthor{awip13} and \citeauthor{kipp14},
including subequent studies by the same group of authors, focused on the
detectability of potentially habitable exomoons, i.e., exomoons located in the
stellar CLI-HZs.  Emphasis has been placed on results from the {\it Kepler}
space mission through the virtue of photodynamics and sophisticated data analysis
techniques. \cite{noyo14} suggested to identify exomoons through observations
of radio emissions; this type of research is guided by the physical structure
of the Jupiter--Io system.

A crucial aspect for the existence of exomoons concerns the necessity of
orbital stability.  Because of three-body interaction, moons will be
lost from their host planet if the distance between the planet and moon
is too large.  This underlying orbital stability limit cannot be calculated
in a straightforward manner, see, e.g., \cite{musi14} and references therein,
although provisional insight can be obtained if assumed that the moon's orbit
must remain within the Hill sphere (or Hill radius) given as
\begin{equation}
R_{\rm H} \ = \ a_p ( 1-e_p ) \Big( \frac{M_p}{3 M_\ast} \Big)^{\frac{1}{3}}
\ \ ;
\end{equation}
see \cite{barn02} and references therein.  Here $M_p$ and $M_\ast$ denote the
mass of the planet and star, whereas $a_p$ and $e_p$ denote the semi-major axis
and eccentricity of the exoplanetary orbit, respectively.

Results are given in Table~5.  If the unprojected planetary mass is
unavailable, ${\sin}i$ is assumed as $\pi/4$ to calculate $M_p$.
For HD~169830c and $\upsilon$~And~Ad, the Hill radius is found as
$R_{\rm H} \simeq 0.25$~AU (see Fig.~7), which allow ample space for
exomoons if successfully formed.  Note that the extent of the Hill radius
is adversely impacted by the exoplanet's orbital eccentricity.  Therefore,
for tutorial reasons, we also convey ${\tilde R}_{\rm H}$ given as
$R_{\rm H} / (1-e_p)$.  Moreover, we also list $R_{\rm H}/a_{\rm p}$
to indicate the relative importance of the Hill radius for each system.
In this regard, we found that the relative extent of the Hill radius is most
pronounced for 30~Ari~Bb, closely followed by $\upsilon$~And~Ad, and least
pronounced for HD~8673b.  However, though widely used, the Hill radius is a
far-from-perfect measure for assessing or predicting the orbital stability
of possible exomoons in those systems as typically only a moderate fraction
of $R_{\rm H}$ is available.  Limitations and modifications of the concept
of Hill radius, also referred to as Hill stability, have been explored by,
e.g., \cite{domi06}, \cite{cunt09}, \cite{hink13}, and others.


\section{Results and discussion}

As part of our study, we explore the impact of the UV levels that
possible exomoons of the eight systems would experience by computing
$E_{\rm eff}$ for the CLI-HZs and for the various exoplanetary orbits.
$E_{\rm eff}$ describes the ratio of UV infliction on DNA for a particular
set of conditions to the case without any atmospheric interference for
an object at 1 AU from the Sun.  Hence, $E_{\rm eff} = 1$ implies that
the set of conditions creates a similar level of UV-based damage as
it would be found immediately above the Earth's atmosphere.  Here we
simulate the case of no atmosphere as well as four cases of different
atmospheric attenuations for each planetary system.

The results of the case without atmospheric attenuation, i.e., ATT$=1$,
are given in Fig.~8. The ranges of $E_{\rm eff}$ corresponding to the
CLI-HZs and to the orbits of possible exomoons (represented by their hosts,
the Jupiter-type planets) are shown by different colors.  As expected, the
UV levels in the CLI-HZs of the F-type stars are generally more severe
than for the solar environment, except for regions beyond the outer limits
of the GHZs.  In the most severe case, found at the inner limit of the EHZ
for HD~8673, $E_{\rm eff}$ is 5.2.  For this star, $E_{\rm eff}$ at the
outer limits of the GHZ and EHZ are 1.1 and 0.53, respectively.  At the
inner limit of the EHZ of HD~153950, which has the lowest $T_{\rm eff}$
of our target stars, $E_{\rm eff}$ is given as 3.2. For HD~153950,
$E_{\rm eff}$ at the outer limits of the CHZ, GHZ, and EHZ are identified as
1.0, 0.63 and 0.32, respectively.

The distances between the host stars
and the various limits of CLI-HZs are strongly correlated with the stellar
luminosities rather than the stellar effective temperatures.  However,
$E_{\rm eff}$ at the limits of the CLI-HZs is mostly correlated with the
star's effective temperature.  For example, $E_{\rm eff}$ at the inner limit
of the EHZ for a relatively hot star is higher than that for a relatively
cool star.  Moreover, the differences between $E_{\rm eff}$ at the various
outer and inner limits of a CLI-HZ are correlated with the stellar effective
temperatures as well, whereas the widths of CLI-HZs mostly depend on the
stellar luminosities.  Thus, $E_{\rm eff}$ shows a steep rate of change
with location within the CLI-HZ for a hot, but less luminous star.

In the case of HD~33564, the EHZ extends from 0.95 to 2.99~AU (see Table~2),
and the values for the $E_{\rm eff}$ at the limits are given as 4.7 and 0.47,
respectively.  In contrast, HD~86264, which is cooler but more luminous than
HD~33564, has a wider EHZ extending from 1.60 to 5.06~AU; thus, the change
in $E_{\rm eff}$ with distance in the CLI-HZ is more gradual.  For HD~86264,
$E_{\rm eff}$ at the inner and outer limit of the EHZ is identified as 4.5
and 0.45, respectively.  However, a wider CLI-HZ does not necessarily imply
that there is a wider area of a relatively mild UV environment.  For example,
the EHZ of HD~8673 extends from 1.39 to 4.37~AU, but the region in the EHZ
with $E_{\rm eff}\leq1$ only extends from 3.17 to 4.37~AU.  HD~15395 has a
narrower EHZ, extending from 1.10 to 3.48~AU, but has a slightly wider region
within the EHZ with $E_{\rm eff}\leq1$; this latter region extends from 1.96
to 3.48~AU.  Also note that for any star considered in our study, the
$E_{\rm eff}$ values obtained at the outer limits of the EHZs are fairly
similar.

The UV environments along the orbits of possible exomoons, hosted by
exoplanets of very high eccentricities, i.e., HD~8673b and HD~86264b (see
Table~4), are most severe, as expected.  Hypothetical exomoons of HD~8673b
and HD~86264b would experience $E_{\rm eff}$ values as high as 14.3 and 15.7,
respectively, at the periapsis; however, these values would decreases to
0.37 and 0.49, respectively, at the exoplanets' apoapsis positions.
Fortunately, following Kepler's second law, excursions of these exoplanets,
and by implication, any possible exomoon, close to the host stars would be
only of relatively short durations considering the relatively high orbital
speeds at those positions.  In contrast, HD~25171b's orbit has a relatively
low eccentricity, entailing that any possible exomoon would face an environment
even milder than that of today's Earth.  Furthermore, any possible exomoon
would be subjected to small fluctuations regarding possible DNA damage along
its orbit, ranging from $E_{\rm eff}=0.38$ to 0.52.  The UV environments for
HD~196830c and $\upsilon$~And~Ad are relatively moderate as well. The
values of $E_{\rm eff}$ at the periapsis and apoapsis of HD~196830c are
identified as 2.1 and 0.52, whereas for $\upsilon$~And~Ad, they are given
as 2.8 and 0.82, respectively.

We also studied the impact of atmospheric attenuation regarding UV
habitability of possible exomoons.  It is found that $E_{\rm eff}$
is greatly reduced for the four cases of atmospheric attenuation taken
into account, with the combinations of attenuation parameters given
as (i)~$(A, B, C)=(0.02,250,0.5)$, (ii)~$(0.05,250,0.5)$,
(iii)~$(0.02,300,0.5)$, and (iv)~$(0.05,300,0.5)$ (see also Paper~I).
Case (iv) attenuates UV radiation most effectively. Case (ii)
is the least effective of the four cases. In Case (iv), $E_{\rm eff}$
in the CLI-HZs is almost at the same level for all systems taken into
account.  Note that case (i) approximately corresponds to the
atmospheric setting of the Archean Earth that has been realized
about $\sim$3.8~Gyr ago.

Regarding HD~8673, $E_{\rm eff}$ is given as 0.17 and 0.018 at the
inner and outer limits of the EHZ, respectively.  Similarly, the values
for the $E_{\rm eff}$ at the inner and outer limits of the EHZ of
HD~153950 are given as 0.13 and 0.013, respectively.  Even Case (ii)
shows a drastic reduction in $E_{\rm eff}$ compared to the case without
atmospheric attenuation.  In this case, $E_{\rm eff}$ is given as 1.5
and 0.15 at the inner and outer limits of the EHZ of HD~8673, respectively.
Moreover, $E_{\rm eff}$ is found as 1.0 and 0.10 at the inner and outer
limit of the EHZ of HD~153950, respectively.  The effect of atmospheric
attenuation is highly significant in extreme situations for possible
exomoons, as, e.g., at the periapsis position of HD~86264b.  $E_{\rm eff}$
in that case is reduced to (i)~4.5, (ii)~4.9, (iii)~1.6, and (iv)~0.6.
On the other hand, $E_{\rm eff}$ at the apoapsis position of HD~86264b
is given as (i)~0.14, (ii)~0.15, (iii)~0.049, and (iv)~0.019.
Concerning possible exomoons of HD~25171b, atmospheric attenuation
as assumed results in even gentler UV environments.  At the periapsis and
apoapsis positions of HD~257171b, $E_{\rm eff}$ is given as (i)~0.15 and 0.10,
(ii)~0.16 and 0.12, (iii)~0.052 and 0.038, and (iv)~0.020 and 0.015,
respectively.  Again, the case (i) is aligned to the atmospheric
attenuation of the Archean Earth.


\section{Summary and conclusions}

The aim of this study is to provide a tentative exploration
of the astrobiological significance of selected F-type stars
based on their UV environments, which host Jupiter-type
planets within or close to their CLI-HZs; it is assumed that
these planets may also be hosts to exomoons.  In this context,
we pursue a balanced consideration of the CLI-HZ and UV-HZ for
those systems, with the orbital stability of the possible exomoons
described by the Hill stability criterion as well as by the work of
\cite{domi06}.  The planets have different orbital properties,
noting that their semi-major axes range from 0.995 to 3.02~AU,
and the eccentricities range from 0.08 (almost circular) to 0.72
(highly elliptical).  Furthermore, the host stars have notably
different masses and ages.  The youngest star is 30~Ari~B, an
early main-sequence star, with an age of $0.91{\pm}0.3$~Gyr
\citep{guen09}.  The most advanced stars are HD~25171 and
HD~153950 with ages of $4.0{\pm}1.6$~Gyr \citep{mout11} and
$4.3{\pm}1.0$~Gyr \citep{mout09}.  Although exomoons have not
yet been detected for systems, the possible existence of moons
stems from inspecting the Solar System, noting that Jupiter and
Saturn possess two to five planet-like moons combined, with the
exact number dependent on the adopted standard.  Generally, even
if exomoons turn out to be absent in our systems of study, our
approach will offer a template for tentatively assessing
habitability in other systems with exomoons, if identified. 

We recognize that the discovery of exomoons in F-star systems is
particularly challenging; in fact, from the current point-of-view it
is beyond existing technology.  Generally, observational constraints
on the atmospheres of exoplanets might be accessible through stellar
transits \citep{kalt10}.  In case of exomoons, a hypothetical opportunity
might exist through exomoon transits around luminous, directly imaged
giant planets \citep{hell14c}.  On the other hand, exomoon detections
around F-type stars will be particularly challenging (and reserved to
future endeavors) because of the tiny moon-to-star radius ratio
\citep{sart99,simo07,hell14a}.   Thus, the value of our study should
be considered in a more general context, as it constitutes a particular
theoretical example of (potential) habitability of exomoons, which we
intend to augment to other cases, including giant planets around stars
of later spectral types, which are transitting.

Our approach includes the calculation of updated sizes
for the CLI-HZs following work by \cite{kopp13,kopp14}, and
most importantly, the role of UV due to the stellar photospheric
radiation based on the approach of Paper~I.  Additionally, we
explore the UV-HZ for the various systems, previously introduced
by \cite{bucc06}, which is found to be about 1.5 times as large as
the GHZ, but smaller than the EHZ as adopted here.  Motivated by
earlier studies, as, e.g., the work by \cite{cock99}, DNA is taken
as a proxy for carbon-based macromolecules following the paradigm
that extraterrestrial biology may be most likely based on hydrocarbons;
see, e.g., \cite{kolb15}.  The DNA action spectrum is utilized
to represent the impact of the stellar UV radiation; this allows
us to assess different systems with or without the consideration
of atmospheric attenuation.  However, no calculation of damage
is made for wavelength below 200~nm, mostly because no data for
action spectra for DNA or other relevant biomolecules could be
located for that regime.  Even though the properties of possible
exomoon atmospheres in the mid- and far-UV regime are unknown,
F-star photospheres are expected to be significant emitters
down to $\sim$160~nm or more (depending on their temperature
and activity range), as previously pointed out by, e.g.,
\cite{fran14} and \cite{cunt16}.

Another import aspect concerns the impact of stellar evolution
and the associated stability of the CLI-HZs; both are largely
determined by the stellar mass --- see Paper~I for details.
Most stars of our sample have masses between 1.1 and 1.3~$M_\odot$.
The most massive star is HD~86264 with a mass of about 1.42~M$_\odot$
\citep{fisc09}.  According to Paper~I, this star should be able
to maintain\footnote{This value is given by the preservation
of the half-width of the respective CLI-HZ as both its inner
and outer limit move outward due to the increase of luminosity
caused by stellar main-sequence evolution.} its CHZ and GHZ for
about 2.1 and 2.9 Gyr, respectively; these values are commensurate
to its age.  Stars with masses of 1.1, 1.2, and 1.3~$M_\odot$
(the majority of our target stars) possess CHZs for time\-spans
of 3.45, 3.1, and 2.7~Gyr, respectively, with the duration of
the GHZs increased by a factor of 1.5 to 2.0.

These time\-scales
are notably shorter than for the Sun; however, geochemical
results for Earth by, e.g., \cite{mojz96} and most recently by
\cite{nutm16} based on the analysis of microbial structures have
convincingly shown that the origin of terrestrial life dates back
to 3.8~Gyr (or longer), indicating that life might in principle be
able to start a few hundred millions of years after the formation
of the hosting object.  This implies that even F-type stars should
in general be able to provide habitable environments.  Moreover,
even for stars like HD~25171 and HD~153950, which already have
left the main-sequence, the provision of biospheric environments
might still be possible as planets or moons originally located
outside of the CLI-HZs might subsequently be able to move in,
which is consistent with the cases studied here.  This possibility
is known as cold-start solution for habitability, as pointed out
by \cite{kast93} and others.

The exomoons, if existing, are considered to be in close
proximity to the Jupiter-type planets; therefore, the
orbits of these planets are decisive for assessing the general
prospects of habitability.  The UV exposures of the exomoons
vary according to the changes in the star--planet distances
dictated by the ellipticity of the planetary orbits.  For example,
it is found that the exoplanets HD~169830c and $\upsilon$~And~Ad
stay within the CLI-HZ at all times, if the most extended limits
are assumed.  Furthermore, $\upsilon$~And~Ad also remains in the
UV-HZ all the time; however, the other exoplanets do not.
Moreover, HD~8673b and HD~86264b make significant excursions beyond
both the inner and outer limits of the CLI-HZ even if the most extended
limits are considered.  This behavior may nullify the exomoon's potential
for habitability depending on its atmospheric and geological conditions,
even though limited excursions from the CLI-HZ may be permissible,
especially for objects of thick atmospheres, as argued by, e.g.,
\cite{will02}.  Furthermore, \cite{abe11} explored habitability
of water-limited objects, sometimes referred to as land-worlds
\citep[e.g.,][]{fran03}, and found that they could remain habitable
much closer to the stars, a result relevant for the systems HD~8673,
HD~33564, HD~86264, 30~Ari~B, and HD~153950.

UV habitability in the context of this study has been measured
by $E_{\rm eff}$, defined as the ratio of damage for a given
distance from the star for an object with or without atmospheric
attenuation, as chosen by the model, relative to the damage for
an object at 1~AU from a solar-like star without atmospheric
attenuation.  Inspections of our results show that for the
hypothetical exomoon environments of the stars HD~8673 and
HD~86264, in the absence of atmospheric attenuation, the
highest values for $E_{\rm eff}$ are attained, given as
about 14 and 15, respectively, which apply to the periapsis
positions of the Jupiter-type planets.  Relative decent 
values of 2.1 (maximum value) and 0.5 (persistent value) are
found for the hypothetical exomoon environments of HD~169830
and HD~25171, respectively.  All values of $E_{\rm eff}$,
including the exceptionally high values, are drastically
reduced if atmospheric attenuation is applied, as identified
for Earth, and as suggested for other objects beyond the
Solar System \citep[e.g.,][]{segu03}.

Our study shows that the damage inflicted on DNA exhibit
a large range of values compared to an Earth-type planet
at Earth-equivalent (i.e., homeothermic) positions in the
solar system.  Typically, these values are relatively high,
especially is at periapsis position.  However, appropriate protection
due to atmospheric attenuation can dramatically increase the chances
of providing habitable environments, which indicates another strong
motivation for the continuation of exoplanet and exomoon focused
atmospheric investigations.  Current results have been given by
\cite{seag10}, \cite{rugh13}, \cite{madh14}, among others.
Considering that the detection of exomoons around different
types of stars (including G, K, and M dwarfs as well as regarding
planets in transit) will remain challenging in the foreseeable future,
we view our work also as a template for future investigation of
habitability.  Among other criteria, the joint consideration of the   
stellar CLI-HZ and UV-HZ appears to be of particular importance.


\acknowledgments
This work has been supported by the Department of Physics,
University of Texas at Arlington.  We also appreciate comments
by R. Heller on an early version of the manuscript.


\clearpage



\clearpage


\begin{figure*}
  \centering
  \epsfig{file=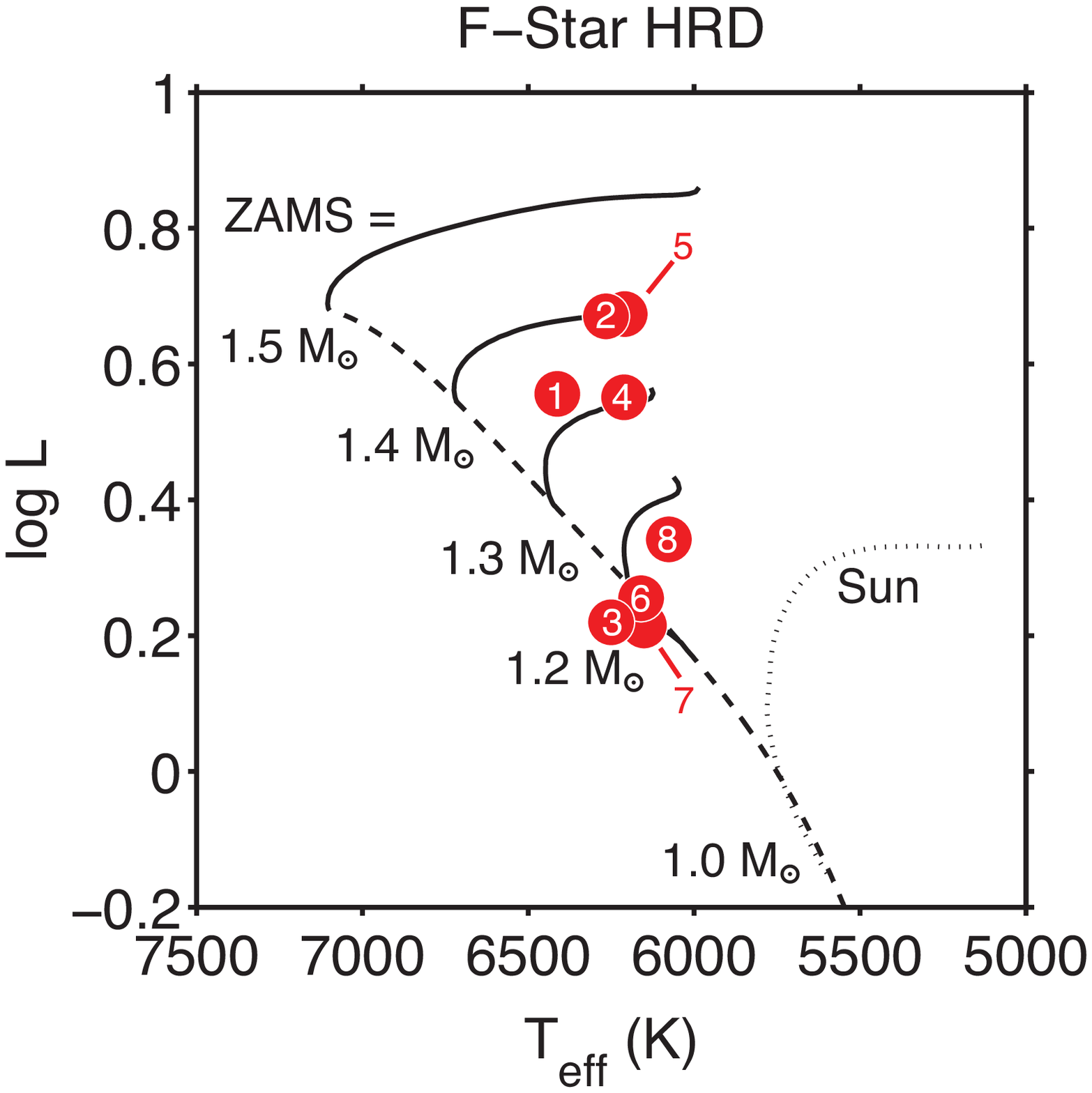,width=1.0\linewidth}
\caption{Stellar evolutionary tracks for F-type stars of different masses
(solid lines) with the main-sequence depicted by a dashed line.
The positions of the eight target stars (see Table~1) considered in
the present study are labeled accordingly.  The evolutionary track
for the Sun (G2~V) is depicted as a dotted line for comparison.
}
\label{fig:1}
\end{figure*}


\clearpage

\begin{figure*}
\centering
  \epsfig{file=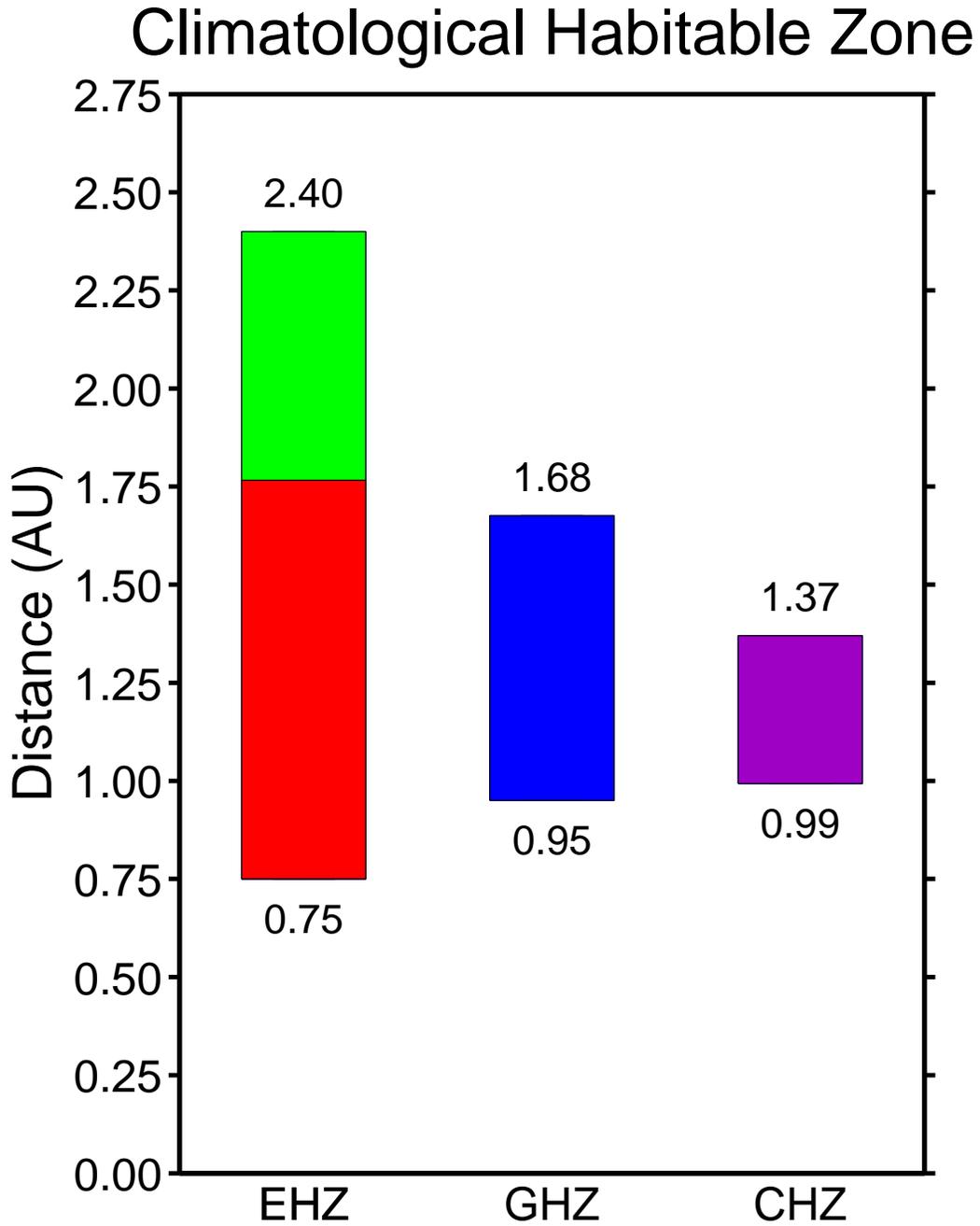,width=0.85\linewidth}
\caption{Types of CLI-HZs; the EHZ, GHZ, and CHZ are color coded.  Regarding the
EHZ, the range defined by the RVEM limits is depicted in red, whereas the
extension based on the work by \cite{misc00} is depicted in green.  
}
\label{fig:2}
\end{figure*}


\clearpage

\begin{figure*}
  \centering
  \epsfig{file=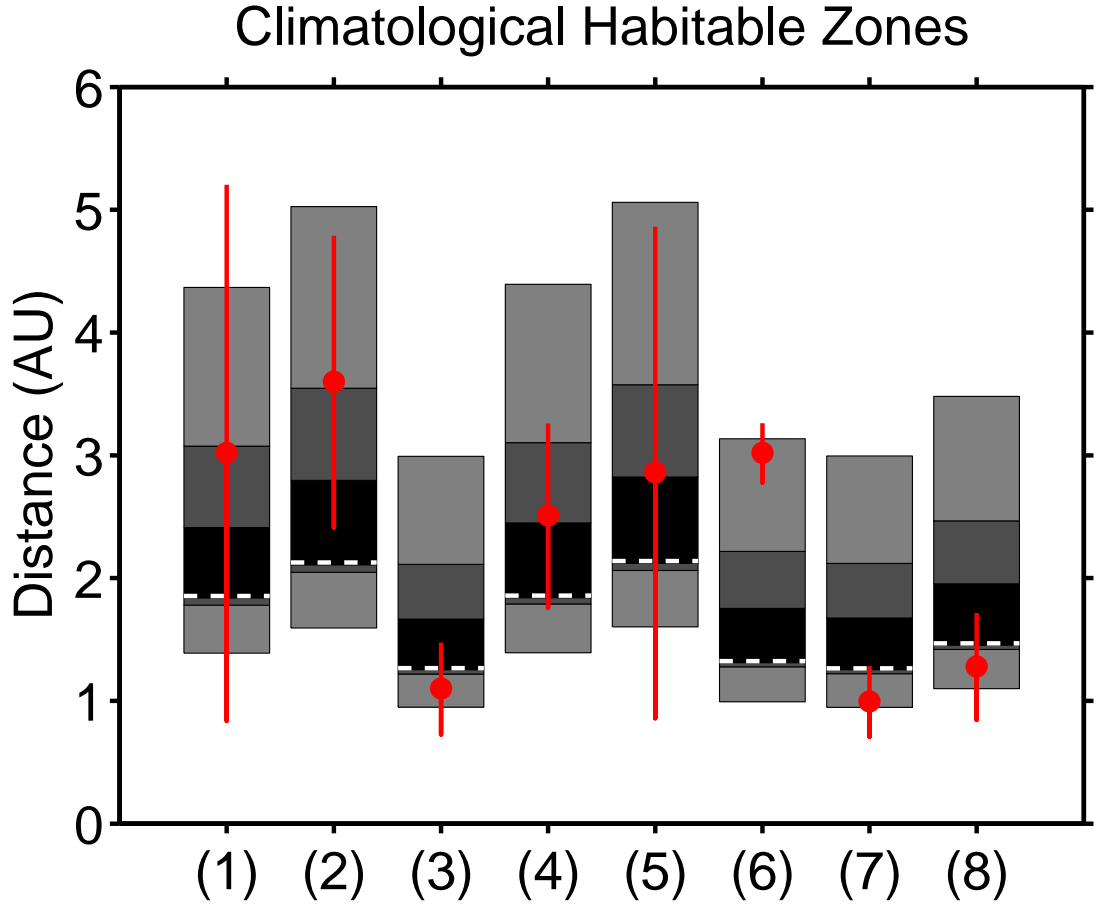,width=1.0\linewidth}
\caption{Domains of the stellar CLI-HZs, subdivided into
the respective CHZs (dark gray), GHZs (medium gray), and EHZs (light gray),
for the various target stars (see Table~1) indicated by (1) to (8).
The Earth-equivalent positions, roughly given by the square root of the
stellar luminosities, are indicated as dashed lines.  The positions of
the giant planets are shown as red dots, whereas the ranges of the
star--planet distances due to the eccentricity of the planetary orbits are
shown as red lines.}
\label{fig:3}
\end{figure*}


\clearpage

\begin{figure*}
  \centering
  \epsfig{file=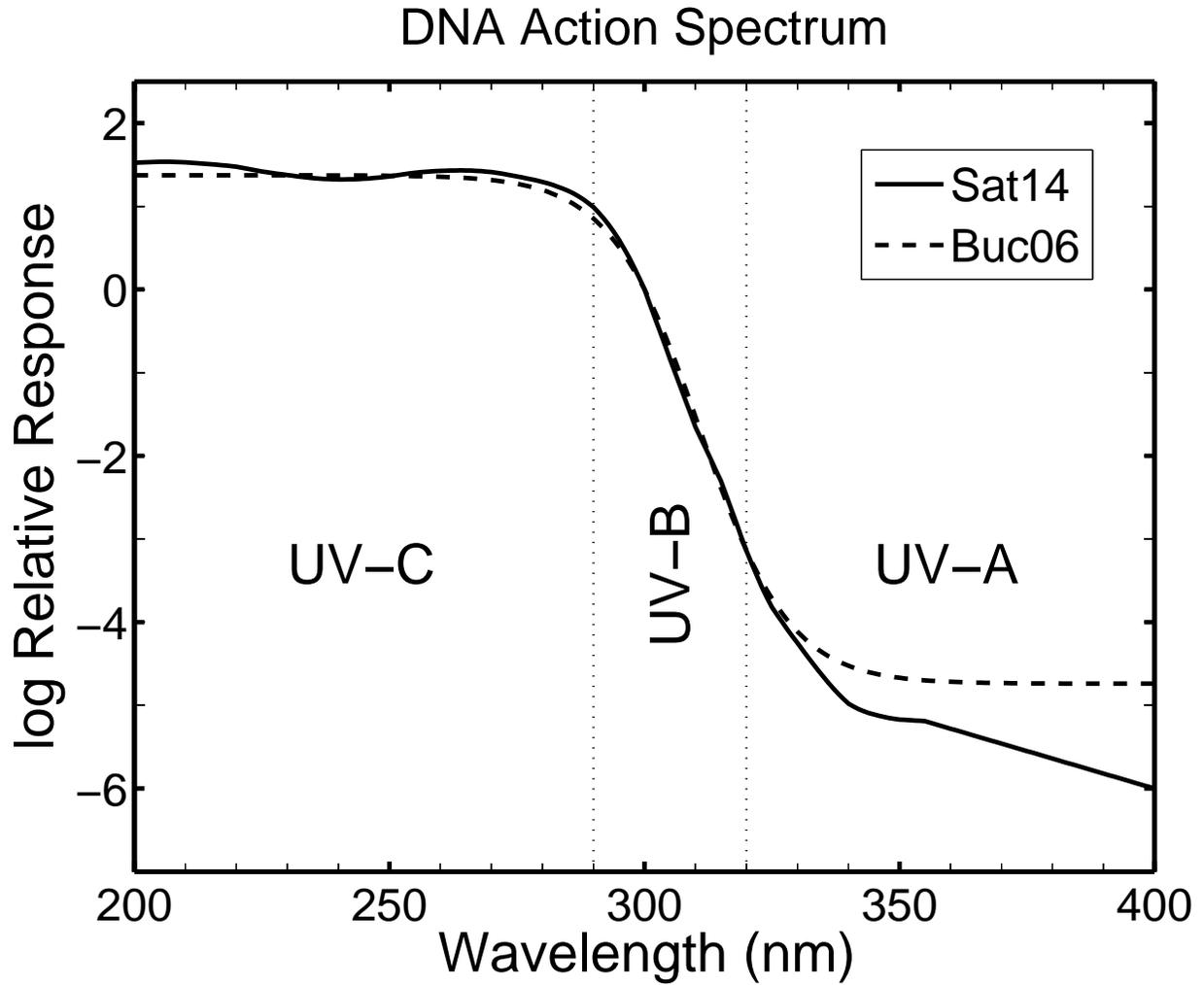,width=1.0\linewidth}
\caption{DNA action spectrum adopted from Paper~I.  The approximation
previously used by \cite{bucc06} is given for comparison.}
\label{fig:4}
\end{figure*}


\clearpage

\begin{figure*}
  \centering
  \epsfig{file=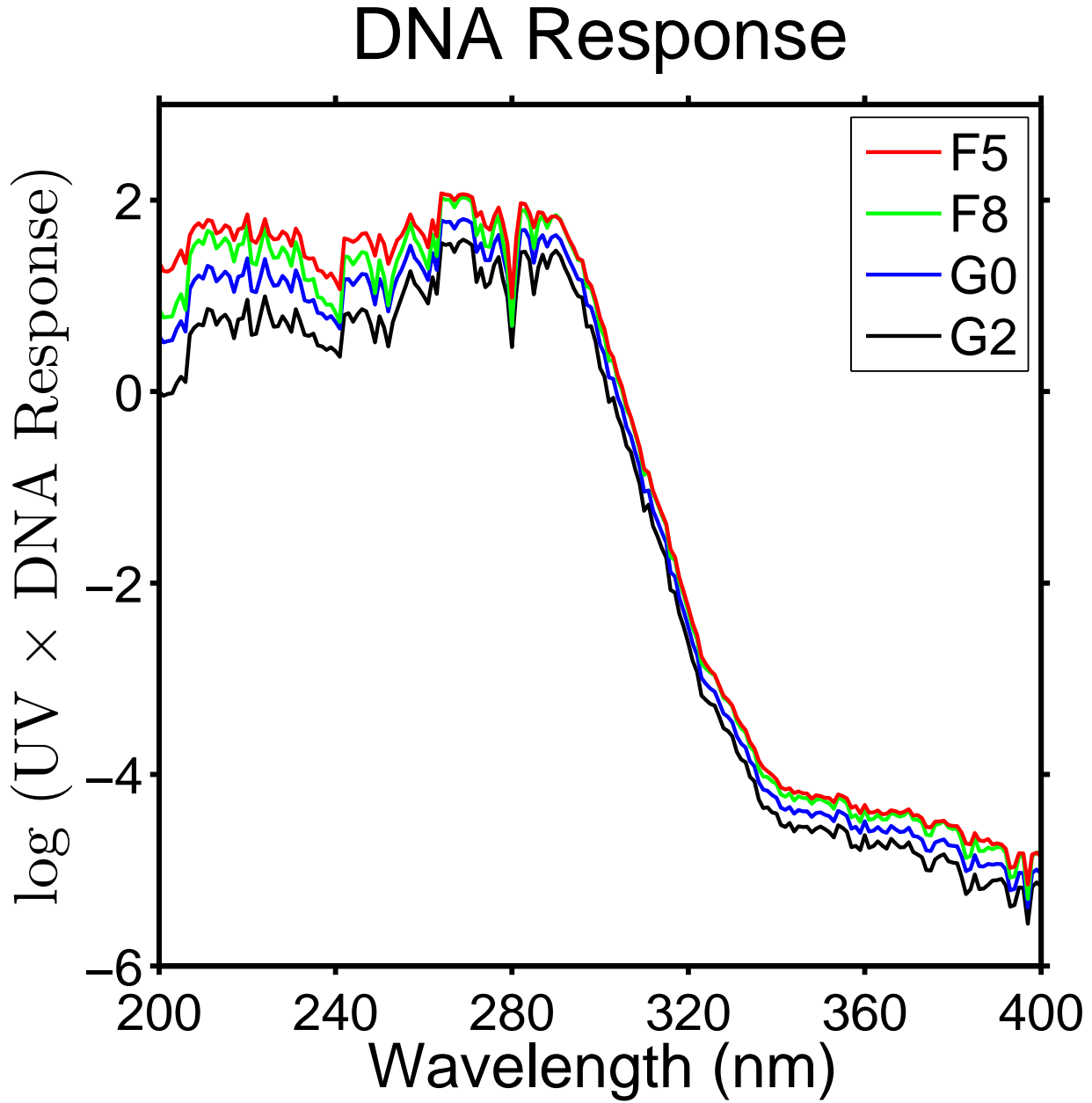,width=1.0\linewidth}
\caption{DNA response between wavelength of 200 and 400~nm pertaining to spectral
irradiation by main-sequence stars between spectral type F5 and G2.  The
stellar spectra have been smoothed assuming a 1~nm running mean bandpass.
Note that the $y$-axis has been normalized to unity for a G2~V star based
on its PHOENIX spectrum.}
\label{fig:5}
\end{figure*}


\clearpage

\begin{figure*}
  \centering
  \epsfig{file=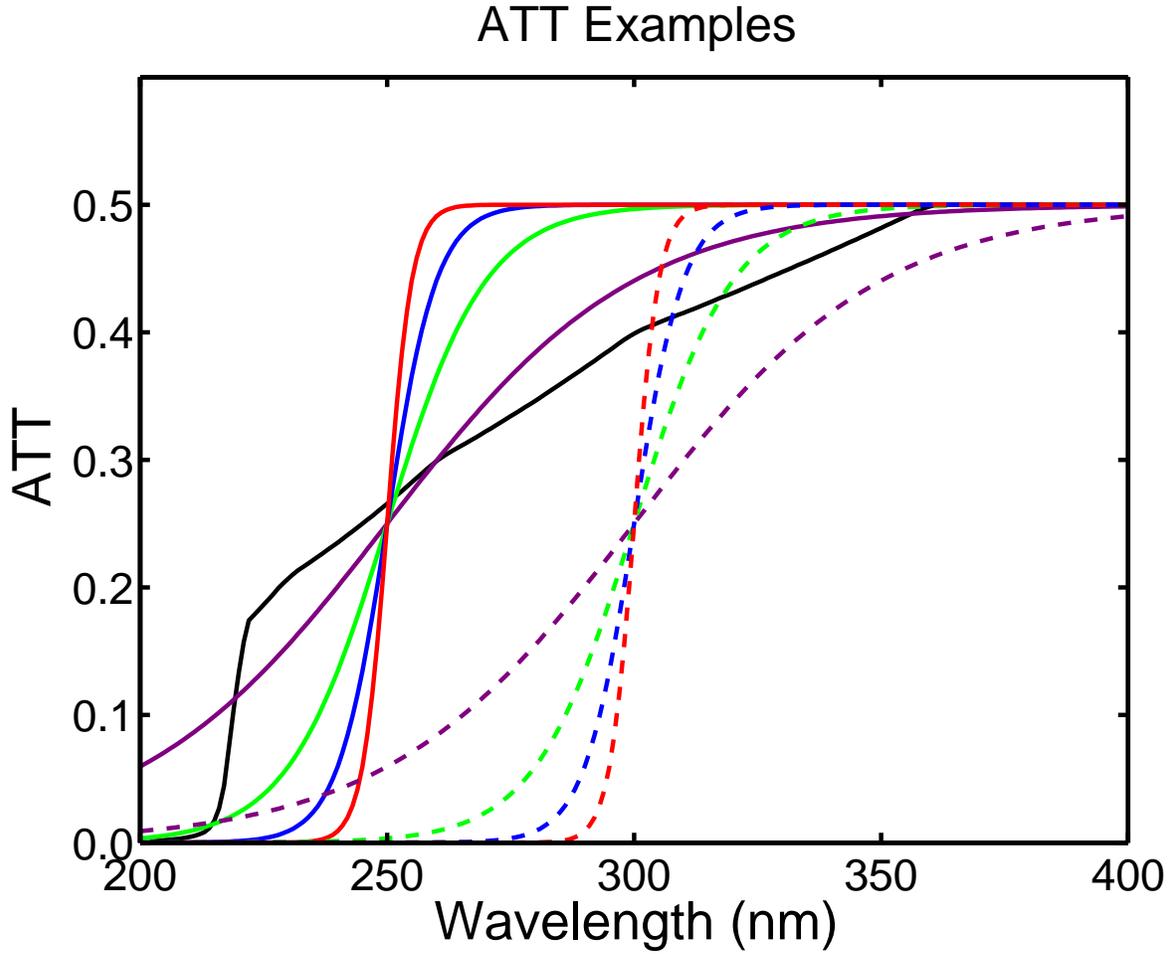,width=0.95\linewidth}
\caption{Examples of parameterized attention functions ATT defined through
$(A, B, C)$ with $C=0.5$ (see Eq.~3).  The solid and dashed colored lines refer
to $B$ = 250 and 300~nm, respectively.  The colors red, blue, green, and purple
refer to $A=$ 0.20, 0.10, 0.05, and 0.02, respectively.  The heavy solid black
line --- that is fairly close to the solid purple line --- indicates the case
as deduced for Earth's Archean eon \citep{cock02}.  Note that the attention
functions are not intended to fit data with the tentative exception of
$(0.02, 250, 0.5)$.}
\label{fig:6}
\end{figure*}


\clearpage

\begin{figure*}
  \centering
  \epsfig{file=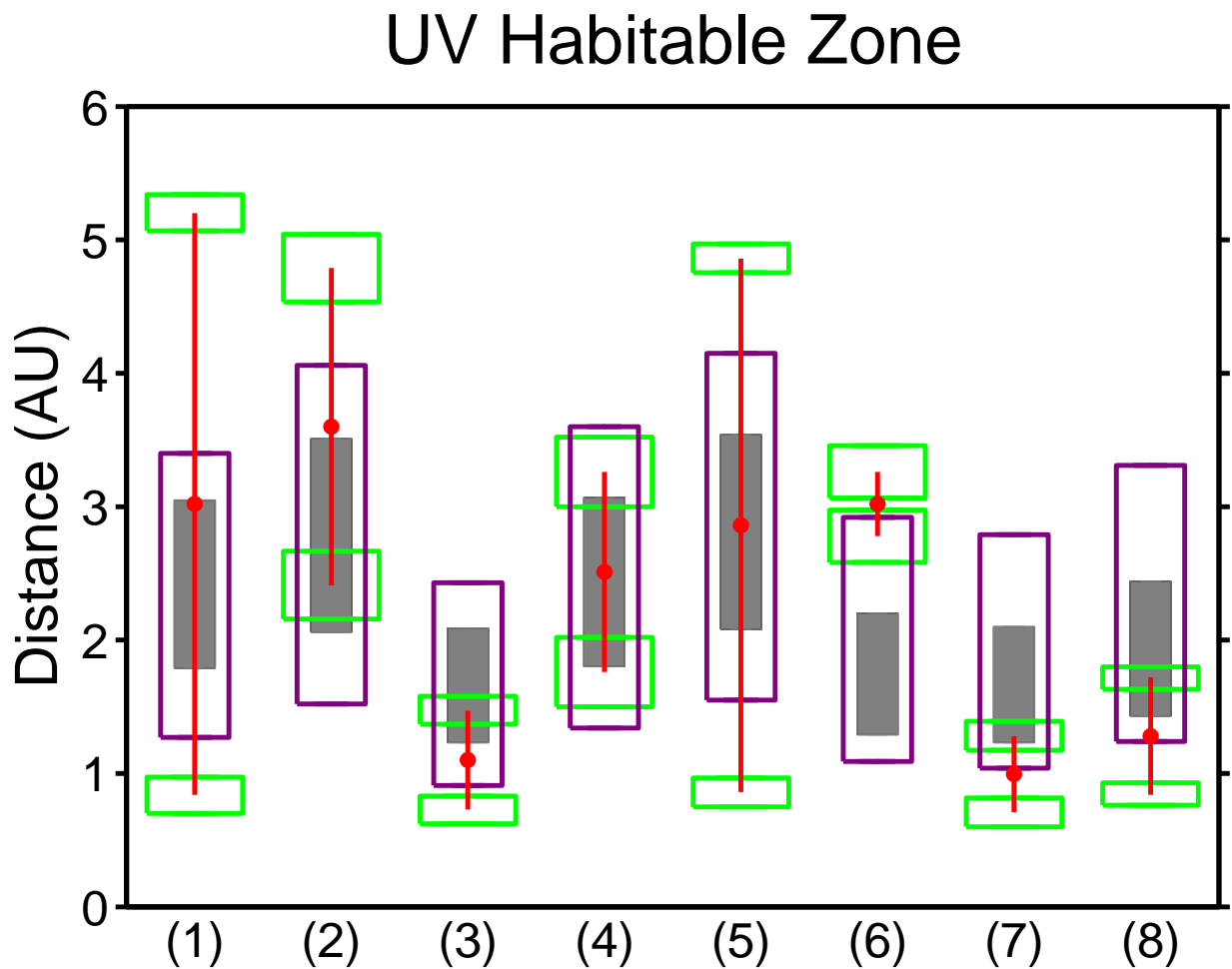,width=1.0\linewidth}
\caption{Depiction of stellar UV-HZs (purple), in comparison to the CLI-HZs, represented
by the GHZs (medium gray), for the various star--planet systems of our study,
indicated by (1) to (8). The positions of the giant planets are shown as red dots,
whereas the ranges of the star--planet distances due to the eccentricity of the
planetary orbits are shown as red lines.  Additionally, for the planetary periapsis
and apoapsis positions, we depict the sizes of the Hill spheres (see Table~5) to convey
the approximate maximal domains of possible exomoons.}
\label{fig:7}
\end{figure*}


\clearpage

\begin{figure*}
  \centering
  \begin{tabular}{c}
  \epsfig{file=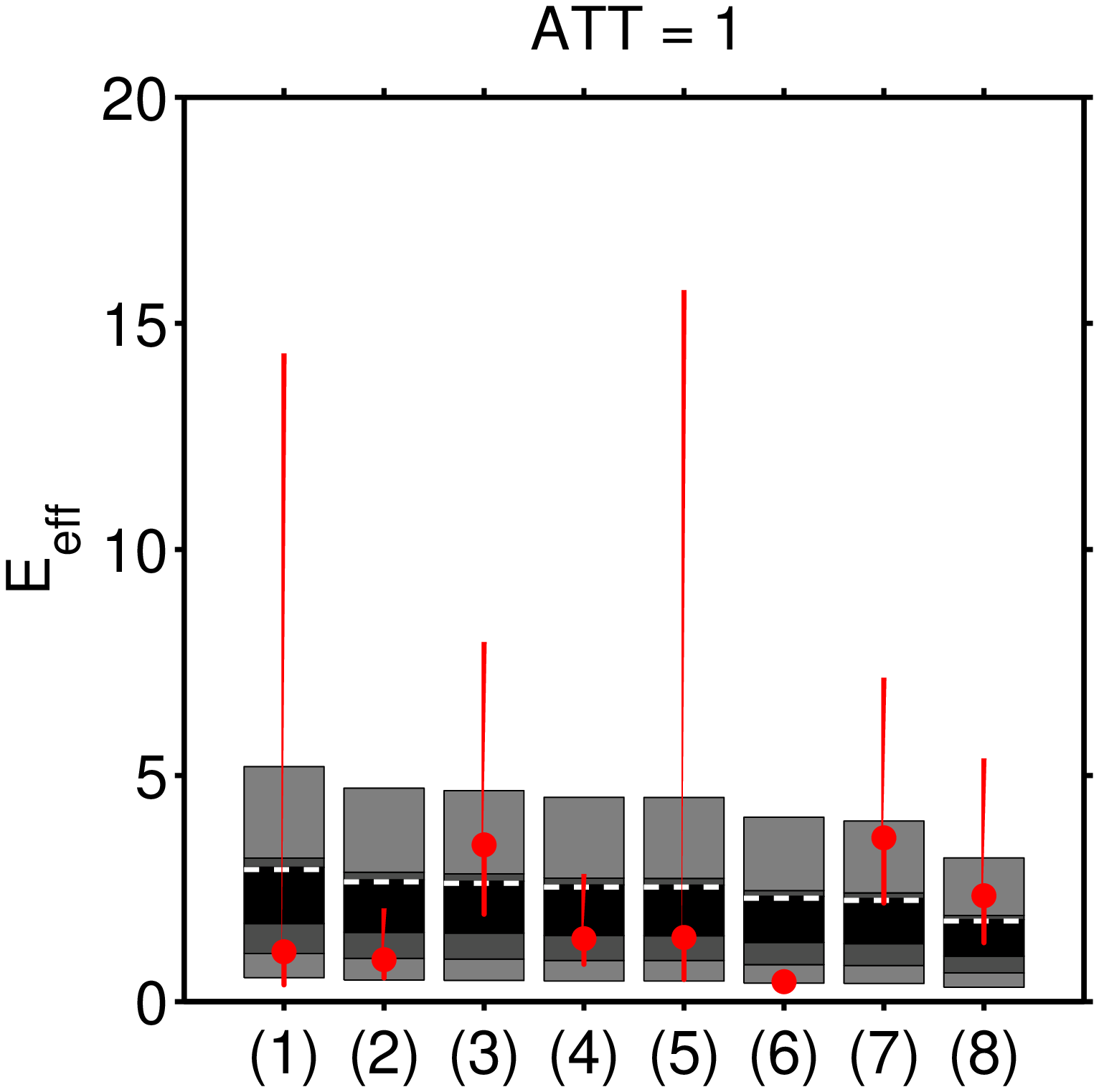,width=0.5\linewidth} \\
  \epsfig{file=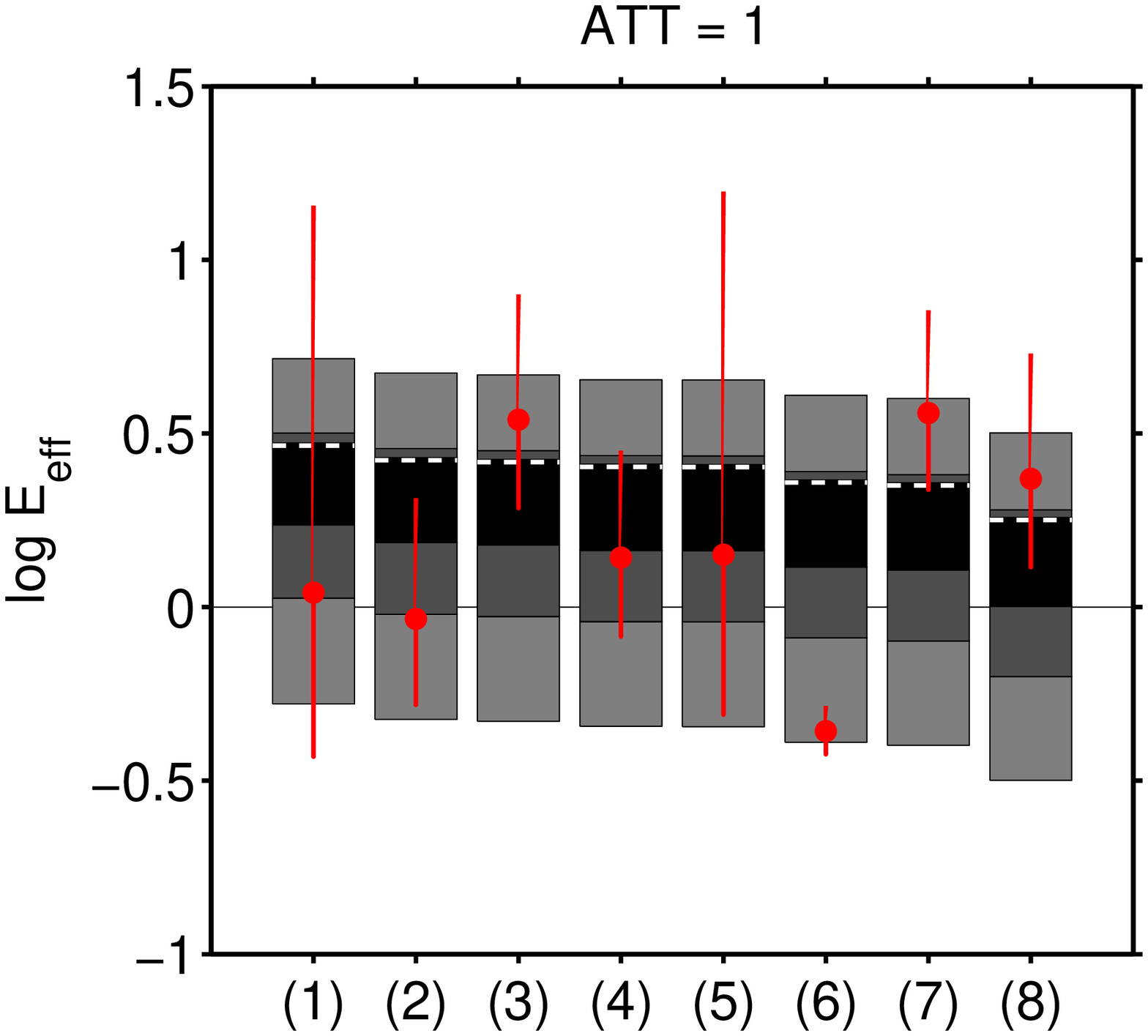,width=0.5\linewidth}
  \end{tabular}
\caption{{\it Top:} $E_{\rm eff}$ for possible exomoons in reference to
the stellar CLI-HZs.  The CHZs, GHZs, and EHZs are indicated by
dark gray, medium gray, and light gray colors.
Earth-equivalent positions within the habitable zones are depicted
by dashed lines.  The target systems (see Table~1) indicated by (1)
to (8).  Additionally, we show the $E_{\rm eff}$ values for
different positions as defined by the Jupiter-type planets.
The ranges in $E_{\rm eff}$, indicated by the red lines, are
due to the eccentric planetary orbits.  The red dots are
depicting $E_{\rm eff}$, when the planetary distance is given
by its semimajor axis.  {\it Bottom:} Replication of the top panel
by choosing a logarithmic scale for enhanced representation.}
\label{fig:8}
\end{figure*}


\clearpage


\begin{deluxetable}{lccccccl}
\tabletypesize{\scriptsize}
\tablecaption{Target Stars}
\tablewidth{0pt}
\tablehead{
Star              & Type   & $T_{\rm eff}$ & $L_\ast$    & $R_\ast$    & $M_\ast$    &  Age    & References \\
...               & ...    & (K)           & ($L_\odot$) & ($R_\odot$) & ($M_\odot$) &  (Gyr)  & ...        
}
\startdata
(1) HD~8673          & F5~V~$-27$~K &	6413 &   3.60 &	1.54  &  1.312 & 2.52 &  1, 2 ($T_{\rm eff}$), 3 ($R_\ast$, $M_\ast$, Age)       \\
(2) HD~169830        & F7~V~$-14$~K &	6266 &   4.69 &	1.84  &  1.4   & 2.3  &  4 ($T_{\rm eff}$, Age), 5 ($R_\ast$, $M_\ast$)          \\
(3) HD~33564         & F7~V~$-30$~K &	6250 &   1.66 &	1.1   &  1.25  & 3.0  &  6 ($T_{\rm eff}$, $M_\ast$, Age), 7 ($R_\ast$)          \\
(4) $\upsilon$~And~A & F8~V~$+12$~K &	6212 &   3.56 &	1.631 &  1.27  & 3.12 &  8 ($T_{\rm eff}$), 9 ($R_\ast$), 10 ($M_\ast$), 3 (Age) \\
(5) HD~86264         & F8~V~$+10$~K &	6210 &   4.72 &	1.88  &  1.42  & 2.24 & 11                                                       \\
(6) HD~25171         & F9~V~$+35$~K &	6160 &   1.80 &	1.18  &  1.09  & 4.0  & 12                                                       \\
(7) 30~Ari~B         & F9~V~$+27$~K &	6152 &   1.64 &	1.13  &  1.11  & 0.91 &  4 ($T_{\rm eff}$, $M_\ast$), 13 ($R_\ast$, Age)                          \\
(8) HD~153950        & G0~V~$+26$~K &	6076 &   2.20 &	1.34  &  1.12  & 4.3  & 14                                                       \\
\enddata
\tablecomments{
For example, an expression such as F5~V $-27$~K means that the stellar effective temperature is 27~K lower
than that of a standard F5~V star noting that the standard $T_{\rm eff}$ values are those of the respective
PHOENIX model.  References:
1: \cite{hart10}, 2: \cite{fuhr08}, 3: \cite{take07},  4: \cite{nord04},  5: \cite{fisc05},  6: \cite{gall05},
7: \cite{pasi01}, 8: \cite{sant04}, 9: \cite{bain08}, 10: \cite{fuhr98}, 11: \cite{fisc09}, 12: \cite{mout11},
13: \cite{guen09}, and 14: \cite{mout09}.  Note that $L_\ast$ are calculated from $T_{\rm eff}$ and $R_\ast$.
}
\end{deluxetable}


\clearpage

\begin{deluxetable}{lccccccc}
\tablecaption{Climatological Habitable Zones}
\tablewidth{0pt}
\tablehead{
System & HZ-iE & HZ-iG & HZ-iC & Earth eqv. & HZ-oC & HZ-oG & HZ-oE \\
...    & (AU)  & (AU)  & (AU)  & (AU)       & (AU)  & (AU)  & (AU)  
}
\startdata
(1) HD~8673           &  1.39 &  1.79 &  1.84 &   1.85     &  2.41 &  3.05 &  4.37  \\
(2) HD~169830         &  1.59 &  2.06 &  2.11 &   2.13     &  2.80 &  3.51 &  5.03  \\
(3) HD~33564          &  0.95 &  1.23 &  1.25 &   1.27     &  1.67 &  2.09 &  2.99  \\
(4) $\upsilon$~And~A  &  1.39 &  1.80 &  1.84 &   1.86     &  2.45 &  3.07 &  4.39  \\
(5) HD~86264          &  1.60 &  2.08 &  2.12 &   2.14     &  2.82 &  3.54 &  5.06  \\
(6) HD~25171          &  0.99 &  1.29 &  1.31 &   1.32     &  1.75 &  2.20 &  3.13  \\
(7) 30~Ari~B          &  0.95 &  1.23 &  1.25 &   1.26     &  1.68 &  2.10 &  3.00  \\
(8) HD~153950         &  1.10 &  1.43 &  1.45 &   1.47     &  1.95 &  2.44 &  3.48  \\
\enddata
\tablecomments{
See Glossary for information on acronyms.  We also give the Earth-equivalent,
i.e., homeothermic, distances.
}
\end{deluxetable}


\clearpage

\begin{deluxetable}{lccc}
\tablecaption{UV Habitable Zones}
\tablewidth{0pt}
\tablehead{
System & HZ-iUV & Earth eqv. & HZ-oUV \\
...    & (AU)   & (AU)       & (AU)
}
\startdata
(1) HD~8673           &  1.27  &  1.85  &  3.40   \\
(2) HD~169830         &  1.52  &  2.13  &  4.06   \\
(3) HD~33564          &  0.91  &  1.27  &  2.43   \\
(4) $\upsilon$~And~A  &  1.34  &  1.86  &  3.60   \\
(5) HD~86264          &  1.55  &  2.14  &  4.15   \\
(6) HD~25171          &  1.09  &  1.32  &  2.92   \\
(7) 30~Ari~B          &  1.04  &  1.26  &  2.79   \\
(8) HD~153950         &  1.24  &  1.47  &  3.31   \\
\enddata
\tablecomments{
See Glossary for information on acronyms.  We also give the Earth-equivalent,
i.e., homeothermic, distances.
}
\end{deluxetable}


\clearpage

\begin{deluxetable}{lcccccl}
\tabletypesize{\scriptsize}
\tablecaption{Planetary Data}
\tablewidth{0pt}
\tablehead{
Planet & $M_{\rm p} {\sin}i$   & $a_{\rm p}$ & $e_{\rm p}$ & $a_{\rm p}(1-e_{\rm p})$ & $a_{\rm p}(1+e_{\rm p})$ & References \\
...    & ($M_{\rm J}$) & (AU)        & ...         & (AU)                     & (AU) & ...                              
}
\startdata
(1) HD~8673b            &  14.2 & 3.02  & 0.723 & 0.84 & 5.20  & \cite{hart10} \\
(2) HD~169830c          &  4.04 & 3.6   & 0.33  & 2.41 & 4.79  & \cite{mayo04} \\
(3) HD~33564b           &  9.1  & 1.1   & 0.34  & 0.73 & 1.47  & \cite{gall05} \\
(4) $\upsilon$~And~Ad   & 10.19 & 2.51  & 0.299 & 1.76 & 3.26  & \cite{curi11} \\
(5) HD~86264b           &  7.0  & 2.86  & 0.7   & 0.86 & 4.86  & \cite{fisc09} \\
(6) HD~25171b           &  0.95 & 3.02  & 0.08  & 2.78 & 3.26  & \cite{mout11} \\
(7) 30~Ari~Bb           &  9.88 & 0.995 & 0.289 & 0.71 & 1.28  & \cite{guen09} \\
(8) HD~153950b          &  2.73 & 1.28  & 0.34  & 0.84 & 1.72  & \cite{mout09} \\
\enddata
\tablecomments{
See main text for original references.  Also, for $\upsilon$~And~Ad, $M_{\rm p}$
is given here instead of the projected value.
}
\end{deluxetable}


\clearpage

\begin{deluxetable}{lccccc}
\tablecaption{Hill Radius}
\tablewidth{0pt}
\tablehead{
Planet & $R_{\rm H}$  & ${\tilde R}_{\rm H}$ & $R_{\rm H}/a_{\rm p}$ \\
...    & (AU)         & (AU)                 & ...                   
}     
\startdata
(1) HD~8673b            &   0.14  &  0.49  &  0.045  \\
(2) HD~169830c          &   0.25  &  0.38  &  0.071  \\
(3) HD~33564b           &   0.10  &  0.16  &  0.095  \\
(4) $\upsilon$~And~Ad   &   0.26  &  0.37  &  0.104  \\
(5) HD~86264b           &   0.11  &  0.36  &  0.038  \\
(6) HD~25171b           &   0.20  &  0.21  &  0.065  \\
(7) 30~Ari~Bb           &   0.11  &  0.15  &  0.109  \\
(8) HD~153950b          &   0.084 &  0.13  &  0.066  \\
\enddata
\end{deluxetable}


\clearpage

\begin{appendix}

\section{Glossary}

\begin{tabular}{ll}
\noalign{\smallskip}
\hline
\noalign{\smallskip}
Acronym & Definition \\
\noalign{\smallskip}
\hline
\noalign{\smallskip}
ArcE        &  Archean Earth                               \\
AU          &  Astronomical Unit                           \\
CHZ         &  Conservative Habitable Zone                 \\
CLI-HZ      &  Climatological Habitable Zone               \\
DNA         &  Deoxyribonucleic Acid                       \\
EHZ         &  Extended Habitable Zone                     \\
GHZ         &  General Habitable Zone                      \\
HZ-iC       &  Conservative Habitable Zone, inner limit    \\
HZ-iE       &  Extended Habitable Zone, inner limit        \\
HZ-iG       &  General Habitable Zone, inner limit         \\
HZ-iUV      &  Ultraviolet Habitable Zone, inner limit     \\
HZ-oC       &  Conservative Habitable Zone, outer limit    \\
HZ-oE       &  Extended Habitable Zone, outer limit        \\
HZ-oG       &  General Habitable Zone, outer limit         \\
HZ-oUV      &  Ultraviolet Habitable Zone, outer limit     \\
RNA         &  Ribonucleic Acid                            \\
RVEM        &  Recent Venus / Early Mars Habitable Zone    \\
UV          &  Ultraviolet                                 \\
UV-A        &  Ultraviolet, A-regime: 400--320 nm          \\
UV-B        &  Ultraviolet, B-regime: 320--290 nm          \\
UV-C        &  Ultraviolet, C-regime: 290--200 nm          \\
UV-HZ       &  Ultraviolet Habitable Zone                  \\
PoM         &  Principle of Mediocrity                     \\
ZAMS        &  Zero-Age Main-Sequence                      \\
\noalign{\smallskip}
\hline
\noalign{\smallskip}
\end{tabular}

\end{appendix}

\end{document}